\documentclass[aps,graphics,twocolumn,showpacs]{revtex4}
\usepackage{graphicx}

\begin{document}
\title{Network implementation of covariant two-qubit quantum operations}
\author{J. Novotn\'y$^{(1)}$, G. Alber$^{(2)}$, I. Jex$^{(1)}$}
\affiliation{$^{(1)}$~Department of Physics, FJFI \v CVUT, B\v rehov\'a 7, 115 19 Praha 1 - Star\'e M\v{e}sto, Czech Republic\\
$^{(2)}$~Institut f\"ur Angewandte Physik, Technische Universit\"at Darmstadt, D-64289 Darmstadt, Germany}
\date{\today}

\begin{abstract}
A six-qubit quantum network consisting of conditional unitary gates is presented
which is capable of implementing a large class of covariant two-qubit quantum
operations. Optimal covariant NOT operations for one and two-qubit systems are special cases contained in this class.
The design of this quantum network exploits basic algebraic properties which also shed new light onto these
covariant quantum processes.
\end{abstract}
 \maketitle

\section{Introduction}

It is well known that certain tasks of information processing cannot be performed
perfectly on the quantum level despite the fact that they can be performed perfectly on a classical
level \cite{Nielsen,Werner0}. Typically,
impossibilities of this kind on the quantum level hint on the existence of corresponding
no-go theorems. They raise interesting questions concerning the optimality
of these quantum processes with respect to particular quality measures. A prominent example in this
respect is the copying of arbitrary quantum states which cannot be achieved perfectly \cite{copy}. The associated problem
of determining quantum operations which can achieve this tasks in the best possible way has stimulated
numerous theoretical and experimental investigations starting with the early work of Bu\v{z}ek and Hillery \cite{copylit}.

Another process of this kind is the quantum NOT transformation
which is to change an arbitrary quantum state into an orthogonal one and which cannot be performed perfectly for
arbitrary input states \cite{one-qubit,Rungta}. Recently, the problem of optimizing quantum NOT processes has been addressed not only for
arbitrary pure one-qubit input states \cite{one-qubit} but also for pure two-qubit input states of a given degree of entanglement \cite{UNOT}.
In this latter context the possible input states are restricted to the set of pure two-qubit states of a given degree of entanglement
which does not constitute a linear vector space. Therefore, the previously
mentioned impossibility arguments concerning quantum NOT operations acting on arbitrary input states  do not apply.
All optimal quantum operations could be determined which perform such a quantum NOT operation
for all possible pure two-qubit input states of a given degree of entanglement with the same quality.
It was demonstrated that these optimal two-qubit quantum NOT operations are members of a convex set of covariant (completely positive) two-qubit
quantum operations. This convex set is generated by four elementary two-qubit quantum operations which form the vertices of a three-dimensional
polytope.
Furthermore, it could be shown that
only in the case of maximally entangled pure two-qubit input states it is possible to perform such a covariant quantum NOT
operation perfectly.
However, so far it is still unknown how this convex set of covariant two-qubit quantum operations
can be implemented in quantum networks with the help of simple elementary quantum gates.

In general, a systematic approach to the problem of designing elementary quantum gate sequences which implement
a given family of covariant quantum operations is not known. In the following it is shown that for the above mentioned
convex set of covariant two-qubit quantum operations
this problem can be solved completely.
This is due to the fact that this convex set of quantum operations has special algebraic properties which can be exploited in a convenient way.
In addition, these algebraic properties shed new light
on the properties of these covariant two-qubit quantum operations.
With the help of additional auxiliary qubits it is possible to design
a quantum network which involves a particular sequence of
conditional unitary qubit gates. Depending on the preparation of the auxiliary qubits any covariant quantum
operation within this convex set can be
implemented by this quantum network.  One of the advantages of this particular network implementation is that the sequence of
conditional unitary qubit gates involved is independent of the covariant quantum operation under consideration.

This paper is organized as follows. In Sec. II basic definitions and properties of the recently introduced
convex set of covariant two-qubit quantum processes \cite{UNOT}
are summarized. The essential algebraic properties of these quantum operations
which are useful for the subsequent construction of the quantum network are discussed in a subsection.
Sec. III addresses the main problem how this convex set of quantum operations can be implemented unitarily by a suitable choice
of auxiliary quantum systems and by an appropriate sequence of elementary quantum gates. As a main result it is shown that any covariant
quantum operation of the convex set discussed in Sec. II can be implemented  by a unitary master transformation which is independent of
the particular quantum operation under consideration.
A quantum network implementation of this main result involving
controlled unitary Pauli operations is discussed in a subsection.

\section{Covariant two-qubit quantum operations}

In this section basic aspects of all completely positive quantum process
are summarized that transform pure two-qubit input states  of a given degree of entanglement in a covariant way.
The recently discussed
optimal quantum NOT operations \cite{UNOT} are special cases thereof.

\subsection{Basic definitions and general properties}
Let us
consider a general completely positive quantum operation $\Pi$ which transforms an arbitrary two-qubit
input state $\rho$ in a covariant way according to
\begin{eqnarray}
\Pi
\left( U_1 \otimes U_2 \rho U_1^{\dagger} \otimes U_2^{\dagger} \right) &=& U_1 \otimes U_2
\Pi(\rho ) U_1^{\dagger} \otimes U_2^{\dagger}.
\label{covariant2}
\end{eqnarray}
Thereby, the requirement of complete positivity ensures that this transformation can be implemented
in a unitary way
possibly with the help of additional auxiliary quantum systems which are
uncorrelated with the two-qubit system initially.
If the covariance  condition (\ref{covariant2})
is satisfied for arbitrary unitary
one-qubit transformations $U_1$,$U_2 \in SU(2)$ \cite{Biedenharn}, it is guaranteed that
the quality of performance of a quantum NOT operation is the same for all possible pure entangled two-qubit
input states of a given degree of entanglement
\cite{UNOT,Werner,Cerf}.

Recently, it was shown \cite{UNOT}
that all possible completely positive covariant two-qubit quantum operations  $\Pi({v,x,y})$ fulfilling Eq.(\ref{covariant2})
form a three-parametric set, i.e.
\begin{equation}
\label{krausform} \rho_{out}= \Pi({v,x,y})\left(\rho \right)=
\sum_{i,j=0}^{3} K_{ij}({v,x,y}) \rho K_{ij}^{\dagger}({v,x,y}),
\end{equation}
with the Kraus operators
\begin{eqnarray}
\label{Krausoperators}
K_{00}({v,x,y})&=&\frac{1}{4}\left(1+3x+3v+9y\right)^{\frac{1}{2}} I
\otimes I,\nonumber\\
K_{i0}({v,x,y}) &=&
\frac{1}{4}\left(1+3x-v-3y\right)^{\frac{1}{2}} \sigma_i \otimes I, \nonumber \\
K_{0i}({v,x,y})&=&\frac{1}{4}\left(1-x+3v-3y\right)^{\frac{1}{2}} I \otimes
\sigma_i, \\
K_{ij}({v,x,y})&=&\frac{1}{4}\left(1-x-v+y\right)^{\frac{1}{2}} \sigma_i
\otimes \sigma_j, \hspace{1em} i,j \in \{ 1,2,3 \},\nonumber
\end{eqnarray}
the unit operator $I$
and the Pauli spin operators $\sigma_1 = X$, $\sigma_2 = Y$, and $\sigma_3 =Z$.
The possible values of the three parameters $x$,$v$ and $y$ are restricted by the requirement of non negativity of
the prefactors entering (\ref{Krausoperators}), i.e.
\begin{eqnarray}
\label{relace_positivity}
1+3x+3v+9y \geq 0, &\hspace{1em}&
1+3x-v-3y \geq 0,\nonumber\\
1-x+3v-3y \geq 0, &\hspace{1em}&
1-x-v+y \geq 0.
\end{eqnarray}
In addition,
trace preservation of the quantum operation  $\Pi({v,x,y})$ implies
\begin{equation}
\sum_{i,j=0}^3 K_{ij}^{\dagger}({v,x,y})K_{ij}({v,x,y})=I.
\end{equation}

An optimal quantum NOT
operation transforms an arbitrary pure two-qubit input state with a given degree of
entanglement
into a not necessarily pure output state of
its orthogonal complement
in an optimal way.
Thereby, the sets $\Omega_{\alpha}$
of pure two-qubit states with a
given degree of entanglement $\alpha\in[0,1/\sqrt{2}]$
are defined by
\cite{Nov1,Bruss}
\begin{eqnarray}
\Omega_{\alpha}=&\Big\{& \big(U_{1} \otimes U_{2}\big) \big( \alpha
|0\rangle \otimes  |0 \rangle +\nonumber\\
&&  \sqrt{1- \alpha^2} |1\rangle \otimes |1\rangle
\big) \Big| U_{1}, U_{2} \in {SU}(2) \Big\} \label{Omega}.
\end{eqnarray}
In the special case $\alpha
=0$ the two-qubit states are separable whereas in the opposite
extreme case $\alpha = 1/\sqrt{2}$ they are maximally entangled.

Let us now summarize some basic properties of such optimal quantum NOT operations \cite{UNOT}:
\begin{itemize}
\item
There is a characteristic threshold value of entanglement at
$\alpha_0=\sqrt{\left(1-\sqrt{1-4K}\right)/{2}}\approx 0.1836$
with $K=\left({8-3\sqrt{6}}\right)/{20}$. For $\alpha \leq \alpha_0$
the optimal quantum NOT operation, i.e. $U_{SEP}$, is independent of the degree of entanglement $\alpha$
and is characterized by the characteristic parameters
$(v=-{1}/{3}, x=-{1}/{3}, y={1}/{9})$ (compare with (\ref{Krausoperators})).
This particular quantum operation is identical to
two optimal covariant one-qubit NOT operations
$u^1$ \cite{one-qubit} applied to each of the input qubits separately, i.e.
$U_{SEP}=u^1 \otimes u^1$ with
\begin{eqnarray}
u^1(\rho)&=&\frac{1}{3}\left( 2I-\rho \right).
\label{u_1}
\end{eqnarray}
These one-qubit NOT operations $u^1$ transform an arbitrary pure one-qubit input state into an orthogonal state in an optimal way
\cite{one-qubit}.
\item
For $\alpha > \alpha_0$
the optimal NOT operations depend on the degree of entanglement $\alpha$ and are characterized by the
parameters (compare with (\ref{Krausoperators}))
\begin{eqnarray}
\label{parametry}
y&=&-\frac{1}{3}\frac{2-31\alpha^2\beta^2+20\alpha^4\beta^4}{-2-35\alpha^2\beta^2+100\alpha^4\beta^4},\nonumber\\
x+v&=&\frac{2}{3}\frac{4-29\alpha^2\beta^2-20\alpha^4\beta^4}{-2-35\alpha^2\beta^2+100\alpha^4\beta^4},\nonumber\\
x,v &\geq& -\frac{1}{3}
\end{eqnarray}
with $\beta=\sqrt{1-\alpha^2}$. \item It can be shown that perfect
NOT operations can be constructed for maximally entangled input
states only. These perfect covariant NOT operations form a
one-parameter family specified by characteristic parameters
fulfilling the conditions $y=-\frac{1}{3}, x+v=\frac{2}{3}$ with
$x,v \geq -\frac{1}{3}$. \item All completely positive covariant
two-qubit processes (\ref{covariant2}) form a three-dimensional
convex set \cite{UNOT}. Any of these processes $\Pi({\bf a})$ can
be represented in the form
\begin{eqnarray}
\label{convex_decomposition} \Pi({\bf a}) &=& a_{00}I +
a_{11}U_{SEP} + a_{01}U_{ME}^{(1)} + a_{10}U_{ME}^{(2)}
\end{eqnarray}
with $a_{mn} \geq 0$ and $\sum_{m,n\in\{0,1\}} a_{mn} =1$.
The quantum operations
\begin{eqnarray}
U_{ME}^{(1)}&=&\Pi(v=1, x=-1/3, y=-1/3),\nonumber\\
U_{ME}^{(2)}&=&\Pi(v=-1/3, x=1, y= -1/3)
\end{eqnarray}
are members of the one-parameter family of
perfect NOT operations for maximally entangled
input states. They are characterized by the additional property that they leave the reduced density operators of
the first ($U_{ME}^{(1)}$) or second ($U_{ME}^{(2)}$) qubit unchanged.
The convex set of quantum processes (\ref{convex_decomposition}) forms a three dimensional polytope whose vertices
are given by the quantum operations
$I$, $U_{SEP}$, $U_{ME}^{(1)}$, and $U_{ME}^{(2)}$. This polytope contains also other interesting quantum operations, such as
the universal two-qubit NOT process
$\mathcal{G}_{NOT}$ studied in Ref.\cite{Rungta}. This latter process is the
optimal NOT operation for all possible pure two-qubit
input states irrespective of their degree of entanglement. Its
convex decomposition is given by
\begin{eqnarray}
\mathcal{G}_{NOT}=0.6~U_{SEP} + 0.2~U_{ME}^{(1)} + 0.2~U_{ME}^{(2)}.
\label{convex1}
\end{eqnarray}
\end{itemize}

\subsection{Algebraic properties}
Let us now explore
further algebraic properties of the covariant two-qubit processes of Eqs.(\ref{covariant2}) and
(\ref{convex_decomposition}).

The vertices
$U_{SEP}$, $U_{ME}^{(1)}$, $U_{ME}^{(2)}$ of the polytope (\ref{convex_decomposition})
are orthogonal and  the operators representing these processes are traceless, i.e.
\begin{eqnarray}
\label{orthogonalita}
Tr\left(U_{SEP}\right)&=&0,~~Tr\left(U_{ME}^{(1)}\right)=0,~~Tr\left(U_{ME}^{(2)}\right)=0,\nonumber\\
Tr\left(U_{ME}^{(1)}U_{SEP}\right)&=&0,~~Tr\left(U_{ME}^{(2)}U_{SEP}\right)=0,\nonumber\\
Tr\left(U_{ME}^{(1)}U_{ME}^{(2)}\right)&=&0.
\end{eqnarray}
Therefore, according to Eq.(\ref{convex_decomposition})
the coefficients ${\bf a} \equiv (a_{00}, a_{01}, a_{10}, a_{11})$ of an
arbitrary covariant two-qubit
quantum operation $\Pi({\bf a})$
are given by
\begin{eqnarray}
a_{00}=\frac{1}{4}Tr\left(\Pi({\bf a})\right),~~a_{11}=\frac{Tr\left(\Pi({\bf a})U_{SEP}\right)}{Tr\left(U_{SEP}^2\right)},\nonumber\\
a_{01}=\frac{Tr\left(\Pi({\bf a})U_{ME}^{(1)}\right)}{Tr\left({U_{ME}^{(1)}}^2\right)},~~
a_{10}=\frac{Tr\left(\Pi({\bf a})U_{ME}^{(2)}\right)}{Tr\left({U_{ME}^{(2)}}^2\right)}.
\end{eqnarray}
Another interesting
feature of the covariant two-qubit quantum operations (\ref{convex_decomposition}) concerns
repeated applications. If two
such quantum operations are applied successively the resulting quantum operation
is again of the form
(\ref{convex_decomposition}). Thus, these quantum operations form a half group.
The coefficients of the convex decompositions of some products of the elementary quantum operations
$U_{SEP}$, $U_{ME}^{(1)}$, and $U_{ME}^{(2)}$
are summarized
in Table \ref{decomposition_multi}. According to this table we have the relation
\begin{equation}
U_{ME}^{(1)}U_{ME}^{(2)}=U_{SEP}.
\end{equation}
Furthermore, it is apparent that the quantum operations $U_{SEP}$,
$U_{ME}^{(1)}$, and $U_{ME}^{(2)}$ commute. Finally let us point
out that the considered covariant processes
(\ref{convex_decomposition}) have nontrivial limit expressions for
$\Pi (a)^n$ for $n\rightarrow \infty$.

\begin{table}[tb]
\renewcommand{\arraystretch}{0.5} {\normalsize
\begin{tabular}{|*{8}{c|}}
\hline \bfseries covariant quantum operation &$~~a_{00}$~~&~~$a_{11}$~~&~~$a_{01}$~~&~~$a_{10}$~~\\
\hline $U_{SEP}^2$ &
${1}/{9}$&${4}/{9}$&${2}/{9}$&${2}/{9}$ \\
\hline
${U_{ME}^{(1)}}^2$ &${1}/{3}$& $0$& ${2}/{3}$& $0$ \\
\hline
${U_{ME}^{(2)}}^2$ & ${1}/{3}$&$0$&$0$&${2}/{3}$ \\
\hline
$U_{SEP}U_{ME}^{(1)} = U_{ME}^{(1)}U_{SEP} $ & $0$&${2}/{3}$&$0$&${1}/{3}$ \\
\hline
$U_{SEP}U_{ME}^{(2)} = U_{ME}^{(2)}U_{SEP} $ & $0$&${2}/{3}$&${1}/{3}$&$0$ \\
\hline
$U_{ME}^{(1)}U_{ME}^{(2)} = U_{ME}^{(2)}U_{ME}^{(1)} $&$0$&$1$&$0$&$0$ \\
\hline
\end{tabular}}
\caption{Convex decompositions of
products of elementary covariant quantum operations which constitute the vertices of the polytope (\ref{convex_decomposition}).}
\label{decomposition_multi}
\end{table}

\section{Quantum network implementation}

In this section it is shown
how an
arbitrary covariant two-qubit quantum operation (\ref{convex_decomposition}) can be implemented in a six-qubit
quantum network by an appropriate
sequence of controlled unitary  gates.
For this purpose it is demonstrated first that
any
covariant two-qubit process (\ref{convex_decomposition}) can be implemented with the help of four auxiliary qubits
by a master
unitary operation. This master unitary operation is independent of the particular covariant
two-qubit quantum operation under consideration.
A particular
covariant two-qubit process is selected by preparing
the
auxiliary four-qubit quantum system in a suitably chosen quantum state.
In a second step a sequence of conditional (unitary) Pauli gates is constructed which implements this unitary master transformation in
this six-qubit quantum network.

\subsection{Unitary representation with auxiliary qubits}
For the purpose of implementing the covariant quantum operations (\ref{convex_decomposition}) unitarily with the help of auxiliary qubits
let us first of all introduce some useful notation. In addition to the four dimensional
Hilbert space ${\cal H}$ of the two-qubit input states we introduce four auxiliary qubits whose
Hilbert space ${\cal H}_{ancilla}$ is sixteen dimensional. The quantum states
$|ijkl\rangle$ with $i,j,k,l\in{0,1}$
are assumed to form an
orthonormal basis in this latter Hilbert space.
We start from the observation that apart from normalization factors
the Kraus operators of (\ref{Krausoperators}) are unitary. Therefore, it is convenient to introduce the corresponding sixteen renormalized
unitary two-qubit operators
\begin{eqnarray}
F_{2i+j~2k+l} &=& \sigma_{2i+j} \otimes\sigma_{2k+l}
\end{eqnarray}
with $\sigma_{0} = I$ and $i,j,k,l\in\{0,1\}$. From these latter
unitary two-qubit operators we can construct the unitary master
transformation
\begin{eqnarray}
\mathcal{U} &=& \sum_{i,j,k,l\in\{0,1\}}F_{2i+j~2k+l}\otimes
|ijkl\rangle \langle ijkl| \label{master}
\end{eqnarray}
which operates on all six-qubits of the Hilbert space ${\cal H}\otimes{\cal H}_{ancilla}$.
Let us assume that initially the four auxiliary qubits are prepared in the mixed quantum state
\begin{eqnarray}
\Sigma({\bf a}) &=& \sum_{i,j,k,l\in\{0,1\}} \frac{a_{{\rm sgn}(i+j)~{\rm sgn}(k+l)}}{3^{
{\rm sgn}(i+j) +
{\rm sgn}(k+l)
}
}|ijkl\rangle \langle ijkl|
\label{sigma}
\end{eqnarray}
with the normalization $a_{00} + a_{01} + a_{10} + a_{11} = 1$ and with ${\rm sgn}(x) = x/|x|$ denoting the signum-function
(${\rm sgn}(0) =0$).
Depending on the values of the coefficients
${\bf a} \equiv (a_{00}, a_{01}, a_{10}, a_{11})$ any covariant quantum process $\Pi({\bf a})$ can be implemented unitarily with the help of
the unitary master transformation (\ref{master}) by preparing the auxiliary four qubits in the quantum state (\ref{sigma}) initially
and by disregarding these four auxiliary qubits after the unitary transformation, i.e.
\begin{eqnarray}
&& \left(a_{00}I + a_{01}U_{ME}^{(1)} + a_{10}1U_{ME}^{(2)} + a_{11}U_{SEP}\right)(\rho) \equiv\nonumber\\
&&\Pi({\bf a}) (\rho) =
{\rm Tr}_{{ancilla}}\left\{
\mathcal{U}\rho \otimes \Sigma({\bf a})
\mathcal{U}^{\dagger}\right\}.
\label{unitaryimpl}
\end{eqnarray}
This unitary implementation of the covariant quantum operations (\ref{convex_decomposition})
is a main result of our paper. It can be proved in a straightforward way by inserting Eqs.(\ref{master})
and (\ref{sigma}) into Eq.(\ref{unitaryimpl}).

Before addressing the general problem of implementing an arbitrary quantum operation of the form of Eq.(\ref{unitaryimpl}) by elementary
quantum gates in this six-qubit quantum network
let us consider the unitary implementation of the covariant quantum operation
$\Pi({a_{00} = 0 = a_{10} = a_{11}, a_{01} = 1}) = U_{ME}^{(1)}$ as an example. For this purpose the auxiliary four-qubit quantum system
has to be prepared in the mixed quantum state
$\Sigma(a_{00} = 0 = a_{10} = a_{11}, a_{01} = 1) = ({1}/{3})\left\{|0001\rangle \langle 0001| + |0010\rangle
\langle 0010| + |0011\rangle \langle 0011|\right\}$.
Thus,
Eq.
(\ref{unitaryimpl}) yields
\begin{eqnarray}
\Pi({a_{00} = 0 = a_{10} = a_{11}, a_{01} = 1})&=&
\frac{F_{01}}{\sqrt{3}}\rho \frac{F^{\dagger}_{01}}{\sqrt{3}}+\nonumber\\
\frac{F_{02}}{\sqrt{3}}\rho \frac{F^{\dagger}_{02}}{\sqrt{3}}+
\frac{F_{03}}{\sqrt{3}}\rho \frac{F^{\dagger}_{03}}{\sqrt{3}}
 &=& U_{ME}^{(1)}.
\end{eqnarray}

\subsection{Network implementation with conditional Pauli gates}

\begin{figure}[h]
\begin{picture}(150,100)(0,0)
\put(75,90){\put(0,0){\circle*{6}} \put(0,-3){\line(0,-1){14}}
\put(0,-20){\circle*{6}} \put(0,-23){\line(0,-1){14}}
\put(0,-40){\circle*{6}} \put(0,-43){\line(0,-1){14}}
\put(0,-43){\line(0,-1){14}} \put(0,-60){\circle*{6}}
\put(0,-63){\line(0,-1){8}} \put(-7,-85){\framebox(14,14){U}}
\multiput(-20,0)(0,-20){4}{\line(1,0){17}
\put(3,0){\line(1,0){17}}}  \put(-20,-78){\line(1,0){13}}
\put(7,-78){\line(1,0){11}}}
\end{picture}
\caption{Quantum circuit representation of the elementary controlled unitary operation $C(U)$ which involves
four control and one target qubit.  Thereby, $U$ denotes a unitary operation acting on the single target qubit which is performed
if and only if the control qubits are in state $|1111\rangle_{control}$.}
\label{CCCCU}
\end{figure}
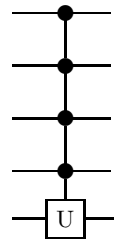

Let us now
implement the unitary master transformation (\ref{master})
by a quantum circuit in the six-qubit quantum network which involves four auxiliary qubits.
According to Eq.(\ref{master})
the quantum circuits have to be designed in such a way that, whenever the four auxiliary qubits are prepared
in a particular quantum state of the computational basis $|ijkl\rangle$ ($i,j,k,l\in\{0,1\}$),
the unitary transformation $F_{2i+j~2k+l}$ is acting onto the two target qubits of the main system with Hilbert space ${\cal H}$.
In order to achieve this goal let us introduce elementary conditional unitary five-qubit quantum gates $C(U)$
which involve four control qubits and one target qubit and whose action on an arbitrary quantum state $|\psi\rangle$ of the target qubit
and a quantum state of the computational basis of the four control qubits $|ijkl\rangle$ is given by
\begin{equation}
\label{def_CCCCU}
C(U)|\psi\rangle_{target}\otimes |ijkl\rangle_{control} =
U^{i\cdot j\cdot k\cdot l}|\psi\rangle_{target}\otimes|ijkl\rangle_{control}
\end{equation}
(compare with Fig. (\ref{CCCCU})).
In other words, the
unitary operation $U$ acts on the target state $|\psi \rangle_{target}$
if and
only if the four control qubits are prepared in the state $|1111\rangle_{control}$.
Universal quantum gates which are capable of implementing such
controlled unitary operations were studied extensively in Ref.
\cite{quantum_gates}, for example.

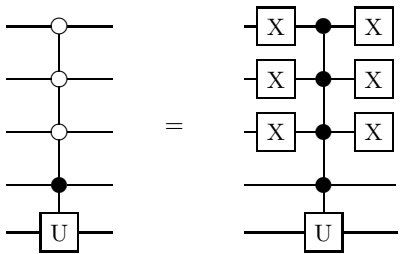
\begin{figure}[h]
\begin{picture}(200,100)(0,0)
\put(50,90){\put(0,0){\circle{6}} \put(0,-3){\line(0,-1){14}}
\put(0,-20){\circle{6}} \put(0,-23){\line(0,-1){14}}
\put(0,-40){\circle{6}} \put(0,-43){\line(0,-1){14}}
\put(0,-43){\line(0,-1){14}} \put(0,-60){\circle*{6}}
\put(0,-63){\line(0,-1){8}} \put(-7,-85){\framebox(14,14){U}}
\multiput(-20,0)(0,-20){4}{\line(1,0){17}
\put(6,0){\line(1,0){17}}}  \put(-20,-78){\line(1,0){13}}
\put(7,-78){\line(1,0){13}}}

\put(90,50){=}

\put(150,90){\put(0,0){\circle*{6}} \put(0,-3){\line(0,-1){14}}
\put(0,-20){\circle*{6}} \put(0,-23){\line(0,-1){14}}
\put(0,-40){\circle*{6}} \put(0,-43){\line(0,-1){14}}
\put(0,-43){\line(0,-1){14}} \put(0,-60){\circle*{6}}
\put(0,-63){\line(0,-1){8}} \put(-7,-85){\framebox(14,14){U}}
\multiput(-30,0)(0,-20){3}{\put(0,0){\line(1,0){5}}
\put(5,-7){\framebox(14,14){X}} \put(19,0){\line(1,0){10}}
\put(32,0){\line(1,0){10}} \put(42,-7){\framebox(14,14){X}}}
\put(-30,-78){\line(1,0){23}} \put(7,-78){\line(1,0){21}}
\put(-30,-60){\line(1,0){27}} \put(3,-60){\line(1,0){24}}}
\end{picture}
\caption{Controlled unitary operation with a unitary operation $U$ acting on
the target qubit
if and only if the first, second, and third control qubits
are in state $|0\rangle$ and the fourth control qubit is in state $|1\rangle$ of the computational basis.}
\label{conditional}
\end{figure}

With the help of the controlled unitary operations
$C(U)$ also other controlled operations can be realized in a straightforward way.
Suppose one wants
to implement a five-qubit quantum gate in which the target qubit is
transformed by a unitary transformation $U$ if and only if the
first, second, and third (control) qubits are in state $|0\rangle$ and the
fourth control qubit is in state $|1\rangle$ of the computational basis.
As apparent from Fig. \ref{conditional} this quantum gate
may be realized
by acting with a Pauli spin operator $X$ onto the control qubits one, two, and three
before and after the application of
the controlled unitary quantum gate $C(U)$.

\begin{figure}[h]
\begin{picture}(200,120)(0,0)
\put(30,110){\put(0,0){\circle{6}} \put(0,-3){\line(0,-1){14}}
\put(0,-20){\circle{6}} \put(0,-23){\line(0,-1){14}}
\put(0,-40){\circle*{6}} \put(0,-43){\line(0,-1){14}}
\put(0,-60){\circle*{6}} \put(0,-63){\line(0,-1){8}}
\put(-7,-85){\framebox(14,14){U}} \put(0,-85){\line(0,-1){8}}
\put(-7,-107){\framebox(14,14){V}}
\multiput(-20,0)(0,-20){4}{\line(1,0){17}
\put(6,0){\line(1,0){17}}}  \put(-20,-78){\line(1,0){13}}
\put(7,-78){\line(1,0){13}} \put(-20,-100){\line(1,0){13}}
\put(7,-100){\line(1,0){13}}}

\put(90,50){=}

\put(150,110){\put(0,0){\circle{6}} \put(0,-3){\line(0,-1){14}}
\put(0,-20){\circle{6}} \put(0,-23){\line(0,-1){14}}
\put(0,-40){\circle*{6}} \put(0,-43){\line(0,-1){14}}
\put(0,-43){\line(0,-1){14}} \put(0,-60){\circle*{6}}
\put(0,-63){\line(0,-1){8}} \put(-7,-85){\framebox(14,14){U}}
\put(0,-85){\line(0,-1){15}}\multiput(-20,0)(0,-20){4}{\line(1,0){17}
\put(6,0){\line(1,0){17}}}  \put(-20,-78){\line(1,0){13}}
\put(7,-78){\line(1,0){13}} \put(-20,-100){\line(1,0){13}}
\put(7,-100){\line(1,0){13}} \put(7,-78){\line(1,0){40}}}

\put(180,110){\put(0,0){\circle{6}} \put(0,-3){\line(0,-1){14}}
\put(0,-20){\circle{6}} \put(0,-23){\line(0,-1){14}}
\put(0,-40){\circle*{6}} \put(0,-43){\line(0,-1){14}}
\put(0,-60){\circle*{6}} \put(0,-63){\line(0,-1){30}}
\put(-7,-107){\framebox(14,14){V}}\multiput(-20,0)(0,-20){4}{\line(1,0){17}
\put(6,0){\line(1,0){17}}}  \put(-20,-78){\line(1,0){13}}
\put(7,-78){\line(1,0){13}} \put(-40,-100){\line(1,0){33}}
\put(7,-100){\line(1,0){13}}}
\end{picture}
\caption{Circuit implementation of a  two-target quantum gate which performs an
operation $U$ on the fifth qubit and an operation $V$ on the sixth
qubit conditional on the first two qubits being in state $|0\rangle$ and qubits three and four being in state $|1\rangle$ of the computational
basis.}
\label{multiple_targets}
\end{figure}
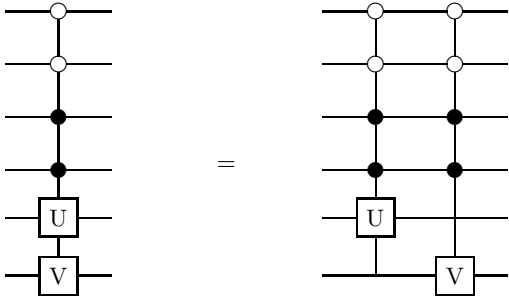

Also multi-target
conditional unitary quantum gates can be realized with the help of the elementary quantum gate $C(U)$.
Such multi-target gates are
natural generalizations of the one-qubit
controlled quantum gates just introduced.
In a general $d$-target conditional unitary quantum gate
a set of unitary operations, say
$\{U_i\}_{i=1}^d$, are performed on $d$ target qubits simultaneously if and only if the control qubits
are prepared in prescribed quantum states.
In Fig.  \ref{multiple_targets} a two-target conditional quantum gate is depicted
in which the unitary operations $U$ and
$V$ are performed on the first and the second target qubit if and only if
the first and the second control qubits are prepared in state $|0\rangle$ and the third and fourth control qubits
are prepared in state $|1\rangle$ of the computational basis.

With the help of such two-target conditional quantum gates a simple sequence
of conditional two-target Pauli gates can
be designed in our six-qubit quantum system which performs the master unitary transformation
(\ref{master}). The circuit scheme of this network is
depicted in Fig. \ref{figure_scheme}. The first four qubits constitute
the control qubits of the auxiliary quantum system.
According to Eq.(\ref{unitaryimpl})
these auxiliary qubits have to be prepared in the quantum state
(\ref{sigma}) initially.
The two input qubits of the main quantum system are prepared in an arbitrary quantum state $\rho$.
The dynamics of the composite
six-qubit quantum system are governed by the master unitary
transformation (\ref{master}) which is implemented by the
network displayed in figure \ref{figure_scheme}.
The action of this dynamics on the two qubits of the main quantum system after having discarded
the four auxiliary qubits is given by the quantum operation (\ref{unitaryimpl}).

\begin{widetext}
\begin{center}
\begin{figure}[h]
\begin{picture}(400,160)(-20,0)
\multiput(-30,130)(0,-20){4}{\put(10,0){\vector(1,0){0}}
\multiput(0,0)(27,0){15}{\put(0,0){\line(1,0){21}}
\put(24,0){\circle{6}}} \put(405,0){\line(1,0){21}}}
\put(-6,130){\put(0,0){\circle*{6}} \put(0,-3){\line(0,-1){14}}
\put(0,-20){\circle{6}} \put(0,-23){\line(0,-1){14}}
\put(0,-40){\circle{6}} \put(0,-43){\line(0,-1){14}}
\put(0,-60){\circle{6}} \put(0,-63){\line(0,-1){30}}
\put(-7,-107){\framebox(14,14){X}} }

\put(21,130){\put(0,0){\circle{6}} \put(0,-3){\line(0,-1){14}}
\put(0,-20){\circle*{6}} \put(0,-23){\line(0,-1){14}}
\put(0,-40){\circle{6}} \put(0,-43){\line(0,-1){14}}
\put(0,-60){\circle{6}} \put(0,-63){\line(0,-1){30}}
\put(-7,-107){\framebox(14,14){Y}} }

\put(48,130){\put(0,0){\circle*{6}} \put(0,-3){\line(0,-1){14}}
\put(0,-20){\circle*{6}} \put(0,-23){\line(0,-1){14}}
\put(0,-40){\circle{6}} \put(0,-43){\line(0,-1){14}}
\put(0,-60){\circle{6}} \put(0,-63){\line(0,-1){30}}
\put(-7,-107){\framebox(14,14){Z}} }

\put(75,130){\put(0,0){\circle{6}} \put(0,-3){\line(0,-1){14}}
\put(0,-20){\circle{6}} \put(0,-23){\line(0,-1){14}}
\put(0,-40){\circle*{6}} \put(0,-43){\line(0,-1){14}}
\put(0,-43){\line(0,-1){14}} \put(0,-60){\circle{6}}
\put(0,-63){\line(0,-1){8}} \put(-7,-85){\framebox(14,14){X}}
\put(0,-85){\line(0,-1){15}}}

\put(102,130){\put(0,0){\circle{6}} \put(0,-3){\line(0,-1){14}}
\put(0,-20){\circle{6}} \put(0,-23){\line(0,-1){14}}
\put(0,-40){\circle{6}} \put(0,-43){\line(0,-1){14}}
\put(0,-43){\line(0,-1){14}} \put(0,-60){\circle*{6}}
\put(0,-63){\line(0,-1){8}} \put(-7,-85){\framebox(14,14){Y}}
\put(0,-85){\line(0,-1){15}}}

\put(129,130){\put(0,0){\circle{6}} \put(0,-3){\line(0,-1){14}}
\put(0,-20){\circle{6}} \put(0,-23){\line(0,-1){14}}
\put(0,-40){\circle*{6}} \put(0,-43){\line(0,-1){14}}
\put(0,-43){\line(0,-1){14}} \put(0,-60){\circle*{6}}
\put(0,-63){\line(0,-1){8}} \put(-7,-85){\framebox(14,14){Z}}
\put(0,-85){\line(0,-1){15}}}


\put(156,130){\put(0,0){\circle*{6}} \put(0,-3){\line(0,-1){14}}
\put(0,-20){\circle{6}} \put(0,-23){\line(0,-1){14}}
\put(0,-40){\circle*{6}} \put(0,-43){\line(0,-1){14}}
\put(0,-60){\circle{6}} \put(0,-63){\line(0,-1){8}}
\put(-7,-85){\framebox(14,14){X}} \put(0,-85){\line(0,-1){8}}
\put(-7,-107){\framebox(14,14){X}} }

\put(183,130){\put(0,0){\circle*{6}} \put(0,-3){\line(0,-1){14}}
\put(0,-20){\circle{6}} \put(0,-23){\line(0,-1){14}}
\put(0,-40){\circle{6}} \put(0,-43){\line(0,-1){14}}
\put(0,-60){\circle*{6}} \put(0,-63){\line(0,-1){8}}
\put(-7,-85){\framebox(14,14){Y}} \put(0,-85){\line(0,-1){8}}
\put(-7,-107){\framebox(14,14){X}} }

\put(210,130){\put(0,0){\circle*{6}} \put(0,-3){\line(0,-1){14}}
\put(0,-20){\circle{6}} \put(0,-23){\line(0,-1){14}}
\put(0,-40){\circle*{6}} \put(0,-43){\line(0,-1){14}}
\put(0,-60){\circle*{6}} \put(0,-63){\line(0,-1){8}}
\put(-7,-85){\framebox(14,14){Z}} \put(0,-85){\line(0,-1){8}}
\put(-7,-107){\framebox(14,14){X}} }

\put(237,130){\put(0,0){\circle{6}} \put(0,-3){\line(0,-1){14}}
\put(0,-20){\circle*{6}} \put(0,-23){\line(0,-1){14}}
\put(0,-40){\circle*{6}} \put(0,-43){\line(0,-1){14}}
\put(0,-60){\circle{6}} \put(0,-63){\line(0,-1){8}}
\put(-7,-85){\framebox(14,14){X}} \put(0,-85){\line(0,-1){8}}
\put(-7,-107){\framebox(14,14){Y}} }

\put(264,130){\put(0,0){\circle{6}} \put(0,-3){\line(0,-1){14}}
\put(0,-20){\circle*{6}} \put(0,-23){\line(0,-1){14}}
\put(0,-40){\circle{6}} \put(0,-43){\line(0,-1){14}}
\put(0,-60){\circle*{6}} \put(0,-63){\line(0,-1){8}}
\put(-7,-85){\framebox(14,14){Y}} \put(0,-85){\line(0,-1){8}}
\put(-7,-107){\framebox(14,14){Y}} }

\put(291,130){\put(0,0){\circle{6}} \put(0,-3){\line(0,-1){14}}
\put(0,-20){\circle*{6}} \put(0,-23){\line(0,-1){14}}
\put(0,-40){\circle*{6}} \put(0,-43){\line(0,-1){14}}
\put(0,-60){\circle*{6}} \put(0,-63){\line(0,-1){8}}
\put(-7,-85){\framebox(14,14){Z}} \put(0,-85){\line(0,-1){8}}
\put(-7,-107){\framebox(14,14){Y}} }

\put(318,130){\put(0,0){\circle*{6}} \put(0,-3){\line(0,-1){14}}
\put(0,-20){\circle*{6}} \put(0,-23){\line(0,-1){14}}
\put(0,-40){\circle*{6}} \put(0,-43){\line(0,-1){14}}
\put(0,-60){\circle{6}} \put(0,-63){\line(0,-1){8}}
\put(-7,-85){\framebox(14,14){X}} \put(0,-85){\line(0,-1){8}}
\put(-7,-107){\framebox(14,14){Z}} }

\put(345,130){\put(0,0){\circle*{6}} \put(0,-3){\line(0,-1){14}}
\put(0,-20){\circle*{6}} \put(0,-23){\line(0,-1){14}}
\put(0,-40){\circle{6}} \put(0,-43){\line(0,-1){14}}
\put(0,-60){\circle*{6}} \put(0,-63){\line(0,-1){8}}
\put(-7,-85){\framebox(14,14){Y}} \put(0,-85){\line(0,-1){8}}
\put(-7,-107){\framebox(14,14){Z}} }

\put(372,130){\put(0,0){\circle*{6}} \put(0,-3){\line(0,-1){14}}
\put(0,-20){\circle*{6}} \put(0,-23){\line(0,-1){14}}
\put(0,-40){\circle*{6}} \put(0,-43){\line(0,-1){14}}
\put(0,-60){\circle*{6}} \put(0,-63){\line(0,-1){8}}
\put(-7,-85){\framebox(14,14){Z}} \put(0,-85){\line(0,-1){8}}
\put(-7,-107){\framebox(14,14){Z}} }

\put(-30,52){\line(1,0){98}}

\multiput(82,52)(27,0){11}{\line(1,0){13}}
\put(379,52){\line(1,0){17}} \put(-30,30){\line(1,0){17}}
\put(1,30){\line(1,0){13}} \put(28,30){\line(1,0){13}}
\put(55,30){\line(1,0){94}}
\multiput(163,30)(27,0){8}{\line(1,0){13}}
\put(379,30){\line(1,0){17}} \put(-20,52){\vector(1,0){0}}
\put(-20,30){\vector(1,0){0}}

\put(-90,90){{\large $\Sigma({\bf a})$}} \put(-55,95){\vector(2,3){22}}
\put(-55,95){\vector(3,2){20}} \put(-55,95){\vector(4,-1){20}}
\put(-55,95){\vector(1,-1){23}} \put(-70,38){{\large $\rho$}}
\put(-55,40){\vector(2,1){20}} \put(-55,40){\vector(2,-1){20}}
\put(415,38){{\large $\Pi({\bf a})(\rho)$}}
\put(400,52){\vector(3,-2){13}} \put(400,30){\vector(1,1){12}}
\end{picture}
\caption{Network consisting of $15$ conditional unitary one- and two-target Pauli gates
which performs the unitary master transformation (\ref{master}) on a six-qubit quantum system.
Initially the four auxiliary qubits are prepared in state $\Sigma({\bf a})$ of (\ref{sigma}) and the two qubits of
the main quantum system are prepared in an arbitrary quantum state $\rho$. After the application of these quantum gates
the two qubits of the main quantum system are in the quantum state $\Pi({\bf a}) (\rho)$ of (\ref{unitaryimpl}).} \label{figure_scheme}
\end{figure}
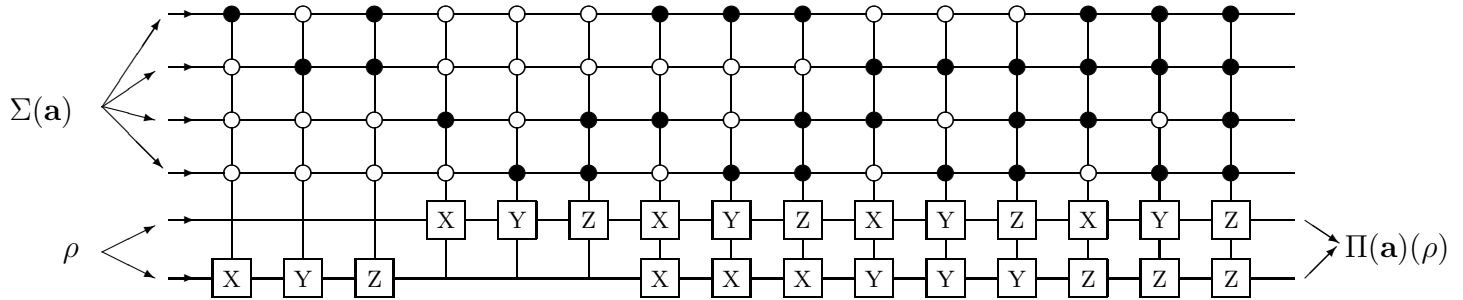
\end{center}
\end{widetext}

\section{Conclusion}
A six-qubit quantum network implementation of all possible
two-qubit quantum operations was presented which transform all
pure two-qubit input states of a given degree of entanglement in a
covariant way. An advantage of this particular implementation is
that it is based on a sequence of conditional Pauli gates which
does not depend on the quantum operation under consideration. A
particular covariant quantum operation is selected by preparing
the four auxiliary qubits in an appropriate quantum state. The
implementation presented rests on special algebraic properties of
these covariant two-qubit quantum operations. Analogous approaches
exploiting similar algebraic properties may also turn out to be
useful for network implementations of other covariant quantum
processes.

\section*{Acknowledgments}
Financial support by GA\v CR 202/04/2101, by the DAAD (GA\v CR
06-01) and by projects LC 06002 (J.N.) and MSM 6840770031 (I.J.)
of the Czech Ministry of Education is acknowledged.


\begin{thebibliography}{99}
\bibitem{Nielsen}
M. A. Nielsen, I. l. Chuang, {\em Quantum Computation and Quantum
Information} (Cambridge University Press, Cambridge, 2000).
\bibitem{Werner0}
A. K. Pati, Phys. Rev. A {\bf 66}, 062319 (2002).
\bibitem{copy} W. K. Wootters and W. H. Zurek, Nature (London) {\bf 299}, 802 (1982).
\bibitem{copylit}  V. Bu\v{z}ek and M. Hillery, Phys. Rev. A {\bf 54}, 1844 (1996).
\bibitem{one-qubit} V. Bu\v{z}ek, M. Hillery, and R. F. Werner, Phys. Rev. A {\bf 60}, R2626 (1999).
\bibitem{Rungta}
P. Rungta, V. Bu\v{z}ek, C. M. Caves, M. Hillery, and G. J.
Milburn, Phys. Rev. A {\bf 64}, 042315 (2001).
\bibitem{UNOT}
J. Novotn\'y, G. Alber, I. Jex, Phys. Rev A 73, 062311 (2006).
\bibitem{Biedenharn}
L. C. Biedenharn and J. D. Louck, {Angular momentum in Quantum
Physics} (Addison-Wesley, Reading, Massachusetts, 1981)
\bibitem{Werner} 
R. F. Werner, Phys. Rev. A {\bf 58}, 1827 (1998); M. Keyl and  R.
F. Werner, J. Math. Phys. {\bf 40}, 3283-3299.
\bibitem{Cerf} 
L.-P. Lamoureux, P. Navez, J. Fiur\'a\v{s}ek, and N. J. Cerf,
Phys. Rev. A {\bf 69}, 040301-1 (2004).
\bibitem{Nov1}
J. Novotn\'y, G. Alber, and I. Jex, Phys. Rev. A {\bf 71}, 042332
(2005).
\bibitem{Bruss}
R. Demkowicz-Dobrzanski, M. Lewenstein, Aditi Sen, Ujiwal Sen, and
D. Bru\ss, Phys. Rev. A {\bf 73}, 032313 (2006).
\bibitem{quantum_gates} A. Barenco, Ch. H. Bennett, R. Cleve,
D. P. DiVincenzo, N. Margolus, P. Shor, T. Sleator, J. Smolin, H.
Weinfurter, Phys. Rev. A {\bf 52}, 3457-3467 (1995).
\end{thebibliography}
\end{document}